\documentclass[%
 reprint,
 superscriptaddress,
 amsmath,amssymb,
 aps,
]{revtex4-2}

\usepackage{graphicx}% Include figure files
\usepackage{dcolumn}% Align table columns on decimal point
\usepackage{bm}% bold math
\usepackage{float}
\usepackage{placeins}
\usepackage{afterpage}

\begin{document}

\preprint{APS/123-QED}

\title{Can Entanglement-enhanced Quantum Kernels Improve Data Classification?}% Force line breaks with \\
%\thanks{corresponding author: dmandal@inst.ac.in}% Force line breaks with \\

\author{Anand Babu}
\affiliation{Quantum Materials and Devices Unit, Institute of Nano Science and Technology, Knowledge City, Sector 81, Mohali 140306, India}

\author{Saurabh G. Ghatnekar}
\affiliation{School of Artificial Intelligence and Data Science, Indian Institute of Technology, Jodhpur, Karwar Jodhpur 342030, India}

\author{Amit Saxena}
\affiliation{Artificial Intelligence \& Quantum Technology Group, Centre for Development of Advanced Computing (C-DAC), Pune, India}

%\author{Dipankar Mandal}
%\affiliation{Quantum Materials and Devices Unit, Institute of Nano Science and Technology, Knowledge City, Sector 81, Mohali 140306, India}

\author{Dipankar Mandal}
\email{Corresponding author: dmandal@inst.ac.in}
\affiliation{Quantum Materials and Devices Unit, Institute of Nano Science and Technology, Knowledge City, Sector 81, Mohali 140306, India}

\date{\today}% It is always \today, today,

\begin{abstract}
Classical machine learning, extensively utilized across diverse domains, faces limitations in speed, efficiency, parallelism, and processing of complex datasets. In contrast, quantum machine learning algorithms offer significant advantages, including exponentially faster computations, enhanced data handling capabilities, inherent parallelism, and improved optimization for complex problems. In this study, we used the entanglement enhanced quantum kernel in quantum support vector machine to train complex respiratory data sets. Compared to classical algorithms, our findings reveal that QSVM performs better with higher accuracy 45\% for complex respiratory data sets while maintaining comparable performance with linear datasets in contrast to their classical counterparts executed on a 2-qubit system. Through our study, we investigate the efficacy of the QSVM-Kernel algorithm in harnessing the enhanced dimensionality of the quantum Hilbert space for effectively training complex datasets.
%\begin{description}
%\end{description}
\end{abstract}

\maketitle

%\tableofcontents

\section{\label{sec:level1}Introduction}

In the ever-evolving landscape of machine learning in various sectors, from accelerating industrial automation to revealing the fundamental aspects of nature, machine learning algorithms have demonstrated remarkable efficacy in processing and analyzing data across multiple dimensions.\cite{Carleo2017, Brunton2021, Babu2023} However, the performance of the ML algorithms is very dependent on the input dataset, having limitations in training random data sets and intricate optimizations. Classical algorithms, such as classical support vector machines (SVM), are extensively utilized in solving various problems in diverse domains; their strength lies in their ability to effectively solve classification problems, particularly through the use of kernel functions; their capability to handle nonlinear relationships between features makes them suitable for a wide range of applications, including bioactivity modeling, protein classification, and image enhancement.\cite{He2019, Kumar2019} As the feature space becomes large and the kernel functions become computationally expensive to estimate, SVM faces challenges in successfully solving such problems. The choice of kernel function, kernel parameter, and regularization parameter are key parameters to effectively training the data sets [5]. Additionally, the computational complexity of increasing the non-linearity of kernels can lead to higher power consumption, posing practical challenges in real-world applications. \cite{Opper2001, Sassi2019}

\begin{figure*}
%\vspace{-1pt} % Adjust this value as needed
\includegraphics[width=\textwidth]{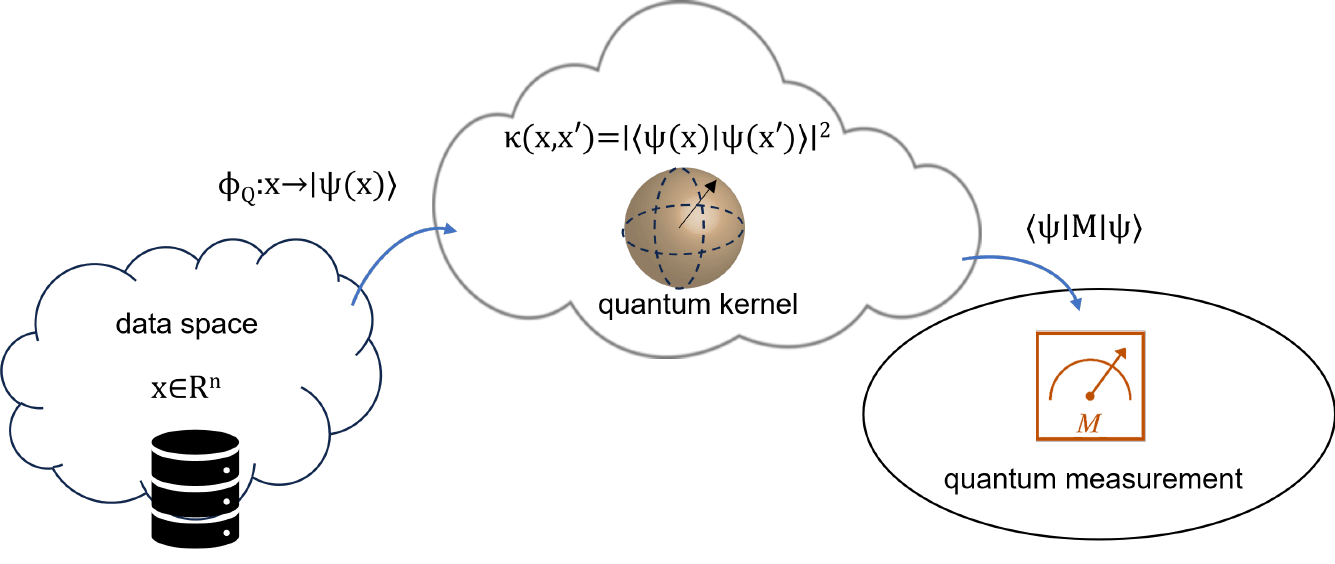}
\caption{(a) Visual representation illustrating the conceptual flow of the Quantum Support Vector Machine (QSVM).}
\vspace{-10pt} % Adjust this value as needed
\end{figure*}

In contrast, quantum machine learning algorithms, including quantum support vector machines, have been performing better in speed, efficiency, and parallel processing of complex datasets compared to their classical counterparts.\cite{Rebentrost2014, Garcia2022, Simoes2023} Different quantum machine algorithms have been utilized for various tasks, including drug discovery \cite{Batra2021}, classification of particles produced by the large hadron collider (LHC) \cite{Wu2021}, and detection of quantum anomalies \cite{Liu2018}, calculation of electronic structure \cite{Xia2018}, and monitoring of healthcare \cite{Flother2023}.  Quantum SVM offers a significant speed-up gain in overall run-time complexity compared to their classical counterparts. \cite{Khan2020} The inherent volatility of random data, its high-dimensional feature spaces, and the absence of clear patterns result in compromised accuracy and computational efficiency. Despite concerted efforts to enhance the performance of classical SVMs in such data sets through custom kernel functions and dimensionality reduction techniques, the problem persists. The ZZ feature map of Quantum Support Vector Machines (QSVM) plays a crucial role in transforming random data into a higher-dimensional space, thereby enhancing the training of QSVM in comparison to classical SVM. It is a non-linear mapping that extracts local properties of the input data, allowing for a more effective representation of the data in a higher-dimensional space. \cite{Schuld2019} This transformation is significant, as it changes the relative position between data points, making the data set easier to classify in the feature space. \cite{Rebentrost2014, Garcia2022} Additionally, the QSVM kernel method utilizes the large dimensionality of the quantum Hilbert space to replace the classical feature space, further enhancing the discriminative power of the QSVM. \cite{Sajjan2022}

In this work, we used QSVM to classify the random dataset of different breathings acquired by the piezoelectric sensor. By merging the principles of quantum computing with the established SVM framework, our approach harnesses the intrinsic parallelism of the quantum realm and the ability to handle superpositions and entanglements. Using quantum-enhanced kernel functions, KQ-SVM seeks to navigate the intricacies of random data distributions and offers a viable solution to classical SVM limitations. Through empirical analyses spanning random-infused datasets, our research validates the superior performance of KQ-SVM, 45\% higher precision than its classical counterparts. Thus, our study makes a pivotal advancement in quantum machine learning, setting a precedent for future explorations into the integration of quantum computing into the realm of data analysis.

\section{\label{sec:level1}Methods}

Kernel methods and quantum computing represent two intriguing yet distinct approaches for deciphering complex data, and while both have their merits, quantum algorithms, particularly quantum Support Vector Machines (SVM), demonstrate superiority, especially when dealing with random datasets. Kernel methods rely on the application of kernel functions to project data into a higher-dimensional feature space, unraveling intricate relationships within the data. This method, while effective, operates within the constraints of classical computation. However, quantum computing leverages the principles of quantum mechanics, utilizing qubits that exhibit superposition and entanglement to manipulate information in ways beyond classical capabilities. \cite{Blank2020, Jager2023} Quantum SVMs, specifically designed for quantum computers, provide a unique advantage by harnessing the power of quantum parallelism to process information more efficiently than classical SVMs. One notable distinction lies in the data representation paradigms employed by these approaches. Kernel methods visualize data as points that reside within the feature space, a representation limited by the classical computational framework. \cite{Park2023, Duan2019} Quantum computers, on the contrary, utilize qubits existing in a vast Hilbert space, allowing for a more nuanced and flexible representation of the data. This fundamental difference underscores the diverse avenues through which information can be captured and manipulated, giving quantum algorithms an edge in handling complex, unpredictable datasets.

Although kernel methods have excelled in various machine learning tasks, boasting a well-established theoretical framework and diverse algorithms, they may face challenges when dealing with highly random datasets where the underlying patterns are elusive and non-linear. Quantum SVMs, on the other hand, offer a promising solution to this issue. The inherent quantum parallelism allows these algorithms to explore multiple solutions simultaneously, providing a more robust approach to capture intricate patterns in seemingly chaotic data. These are computationally demanding problems where classical SVMs and kernel methods may struggle due to their inherent limitations. The quantum advantage lies in its ability to process large amounts of information in parallel, offering a potential breakthrough for solving problems that were once deemed impractical for classical computation.

% \begin{figure}
% \includegraphics[width=8.5cm]{figure_1.pdf}
% \caption{\label{fig:wide}(a) Illustrating the concept of the SVM and quantum SVM. (b) Visual representation illustrating the conceptual flow of the Quantum Support Vector Machine (QSVM).}
% \end{figure}

The captivating journey into the heart of a quantum support vector machine (QSVM) is a meticulous exploration of the intricate dance of quantum states, feature transformations, and learning algorithms that orchestrate this powerful machine learning tool. The journey begins with the preparation of qubits, the fundamental building blocks of quantum computation, in a specific configuration, laying the foundation for subsequent transformations. \cite{Duan2019} The dynamical map then takes center stage, orchestrating the evolution of the quantum state under the combined influence of the input data and the chosen kernel function. \cite{Blank2020, Bennett2000} This map acts as a translator, encoding the complex relationship between raw data and the feature space where classification ultimately occurs. \cite{Rebentrost2014} As the qubits evolve through this map, their state transforms into the evolved density matrix, reflecting the inherent uncertainty that defines the quantum realm. The measured feature vector then collapses the quantum wavefunction, transforming the probabilistic quantum state into a concrete classical vector suitable for classification algorithms. This vector serves as the bridge between the quantum realm and the classical world, carrying the distilled essence of the data within the feature space. The feature map plays a pivotal role in this transformation, acting as a portal that transports the data from its original input space to a higher-dimensional realm known as the feature space. Within this expanded canvas, complex relationships between data points that were previously hidden can become readily apparent, potentially leading to superior classification accuracy in challenging datasets. The training function plays a crucial role in guiding the behavior of the dynamical map and the resulting feature map, ultimately enabling the QSVM to navigate the vast feature space and distinguish between classes effectively. By meticulously optimizing this function through a training process, the QSVM gradually refines its ability to separate the data in the feature space, ultimately leading to more accurate classifications (Figure 1). \cite{Simoes2023, Vasques2023}
Classical SVM seeks to find a hyperplane that maximizes the margin between the two classes. The decision function f(x)  for SVM is

\begin{eqnarray}
f(\mathbf{x}) = \mathbf{w} \cdot \phi(\mathbf{x}) + b
\end{eqnarray}
where $\phi$:Rd→F is the feature map that transforms the input data into a higher-dimensional feature space F. The optimization problem is

\begin{eqnarray}
    \min_{\mathbf{w}, b} \frac{1}{2} \|\mathbf{w}\|^2 + C \sum_{i=1}^{2n} \max(0, 1 - y_i (\mathbf{w} \cdot \phi(\mathbf{x}_i) + b))
\end{eqnarray}

Classical SVM uses a feature map \(\phi: \mathbb{R}^n \to H\) to map input data \(x \in \mathbb{R}^n\) to a higher-dimensional feature space H as $\phi$(x), with the kernel function.

\begin{eqnarray}
    K(x_i, x_j) = \phi(x_i) \cdot \phi(x_j)
\end{eqnarray}

This leads to the following decision function of SVM.
\begin{eqnarray}
   f(\mathbf{x}) = \sum_{i=1}^N \alpha_i y_i K(\mathbf{x}_i, \mathbf{x}) + b
\end{eqnarray}

Here, \(\alpha_i\) are the Lagrange multipliers, \(y_i \in \{-1,1\}\) are the labels, and \(b\) is the bias term.

While quantum feature maps input x to a quantum state |$\phi$q(x)⟩ in Hilbert space Hq
\begin{eqnarray}
|\phi_q(\mathbf{x})\rangle = U(\mathbf{x})|0\rangle
\end{eqnarray}

Entangling gates such as the CNOT gate create correlations between qubits
\begin{eqnarray}
   \text{CNOT}(|0\rangle \otimes |+\rangle) = \frac{1}{\sqrt{2}}(|00\rangle + |11\rangle)
\end{eqnarray}
Where U (x) is a quantum circuit parameterized by x., which leads to generating the quantum kernel as an inner product between quantum states.
\begin{eqnarray}
    K_q(\mathbf{x}_i, \mathbf{x}_j) = |\langle \phi_q(\mathbf{x}_i) | \phi_q(\mathbf{x}_j) \rangle|^2
\end{eqnarray}

Entanglement-Enhanced Quantum Kernel
Quantum feature maps embed data into an exponentially larger space, enabling better separation of complex data distributions:
\begin{eqnarray}
    \mathcal{H} = \text{span}\{|0\rangle, |1\rangle, \ldots, |2^n - 1\rangle\}
\end{eqnarray}

Entangled states represent dependencies between features more effectively than classical methods
\begin{eqnarray}
    |\phi(\mathbf{x})\rangle = \sum_{k=0}^{2^n-1} c_k(\mathbf{x}) |k\rangle
\end{eqnarray}

The quantum kernel naturally incorporates non-linear boundaries, making it ideal for datasets with complex structures
\begin{eqnarray}
    K_{\text{quantum}}(\mathbf{x}_i, \mathbf{x}_j) = \left| \sum_{k=0}^{2^n-1} c_k^*(\mathbf{x}_i) c_k(\mathbf{x}_j) \right|^2
\end{eqnarray}

Further entanglement maps to an entangled quantum state |$\phi$q,e(x)⟩,
\begin{eqnarray}
   |\phi_{q,e}(\mathbf{x})\rangle = U_e(\mathbf{x})|\text{entangled state}\rangle
\end{eqnarray}
Where Ue(x) is an entanglement based on the input x. Further, leads to the entanglement enhanced quantum kernel.
\begin{eqnarray}
K_{q,e}(\mathbf{x}_i, \mathbf{x}_j) = |\langle \phi_{q,e}(\mathbf{x}_i) | \phi_{q,e}(\mathbf{x}_j) \rangle|^2
\end{eqnarray}

\begin{eqnarray}
    f(\mathbf{x}) = \sum_{i=1}^N \alpha_i y_i K_{q,e}(\mathbf{x}_i, \mathbf{x}) + b
\end{eqnarray}

The higher accuracy of the quantum SVM is attributed to the entangled quantum states effectively mapping data to a much higher-dimensional space compared to classical or nonentangled quantum mappings, capturing intricate correlations between features, representing complex patterns more effectively, and the decision function now leverages the enhanced kernel.

\section{\label{sec:level1}Results and discussion}

In our investigation of unfolding the power of kernel-enhanced quantum machine learning (QML) model, such as the Kernel-Enhanced Quantum Support Vector Machine (KQ-SVM), on random datasets compared to classical SVMs, the following equations have been considered: 1. Classical SVM Optimization Problem: The classical SVM solves the following optimization problem to find the optimal hyperplane. In order to test the strength of the KQ-SVM in comparison to the classical SVM, various data sets have been selected, such as the breast cancer data set, the Iris data set and the randomly generated respiratory data sets. Figure 2(a) provides a visual representation of the respiratory dataset in a two-dimensional feature space, where f1 on the x-axis and f2 on the y-axis represent the two features. The breast cancer dataset has been taken as a linear dataset, where the two classes are distinguishable (Figure 2(b)). A quantum circuit of 2 qubits comprising the two Hadamard gates to create the entanglement and two ploy x-gates has been utilized to perform the quantum measurement of both datasets (Figure 2c). The QSVM enhanced with the kernel has been found to perform more accurately with 45 \% higher accuracy for the randomly acquired respiratory dataset while providing almost comparable performance for the separate classes of the breast cancer dataset.

\begin{figure}
%\vspace{-10pt}
\includegraphics[width=\columnwidth]{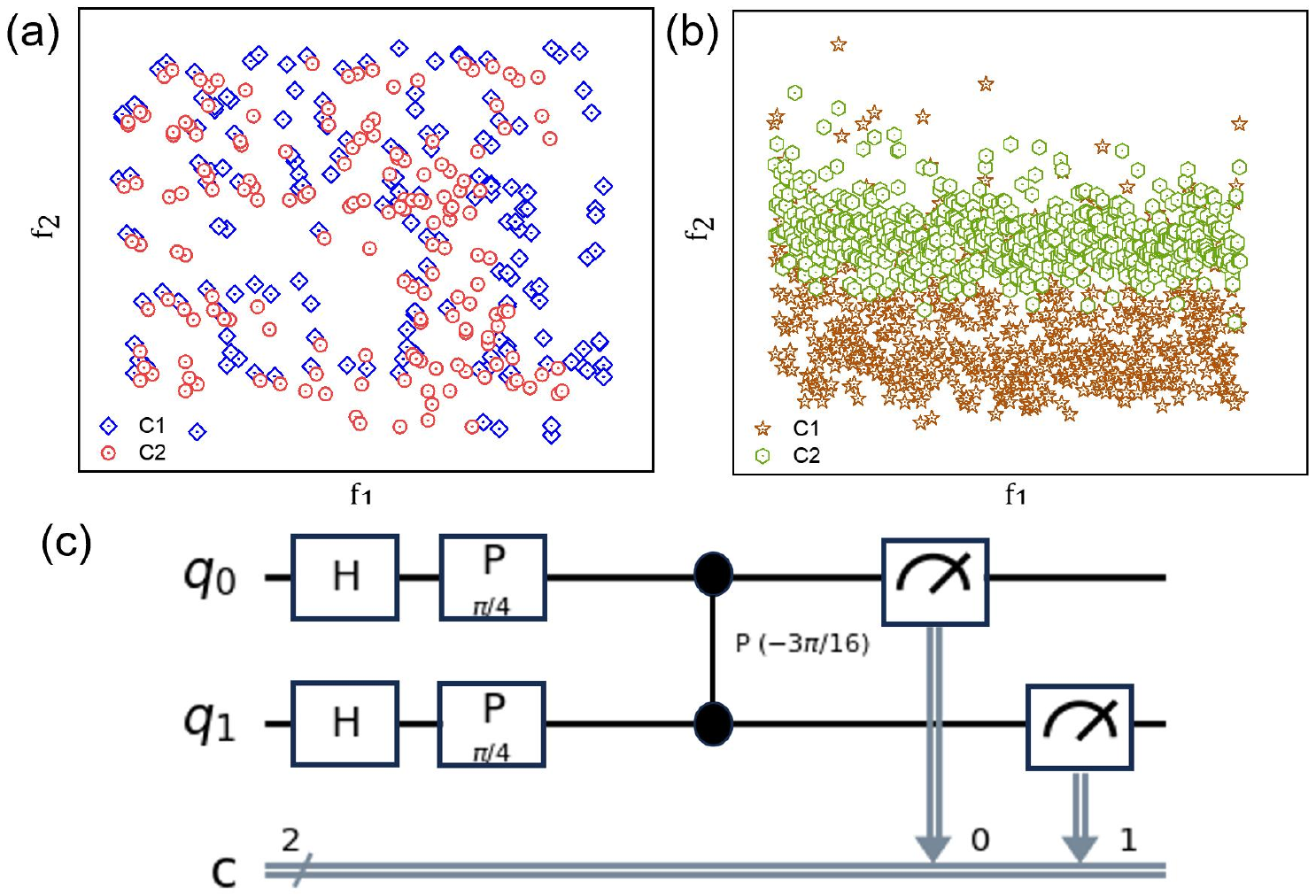}
\caption{(a) Respiratory complex dataset. (b) Breast cancer dataset. (c) 2-qubit equivalent quantum circuit.}
%\vspace{-10pt}
\end{figure}

This approach holds promise for more accurate classifications by addressing complex relationships within the data. \cite{Xu2022} Moreover, depicts the learning journey of a quantum circuit, showing that as the depth of the circuit increases, its training precision increases, indicating its ability to grasp more refined patterns in the data. This suggests that the model effectively uses quantum computing to learn complex relationships and improve its diagnostic capabilities.

\begin{figure}
\includegraphics[width=8.5cm]{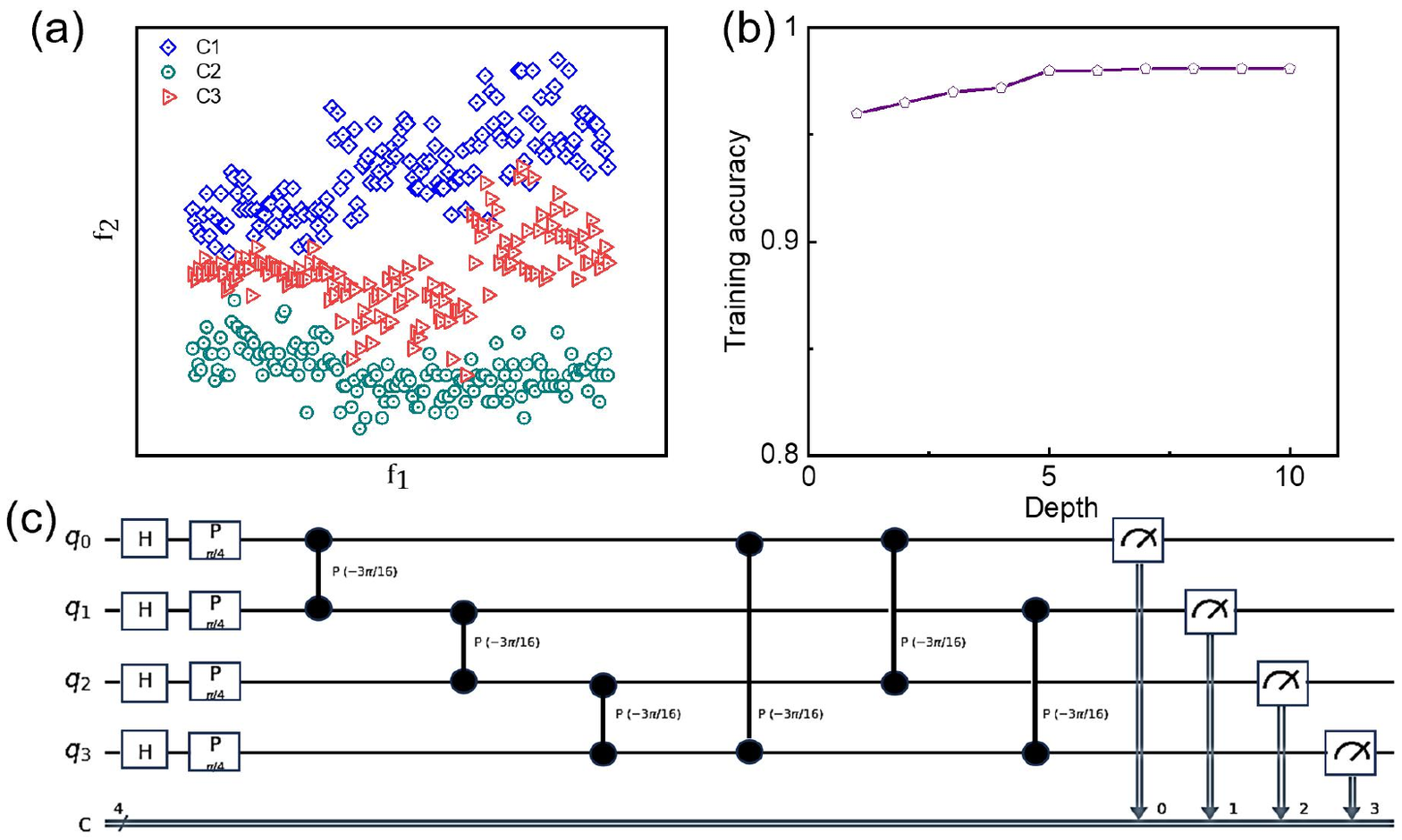}
\caption{(a) Iris dataset for different classes. (b) Training accuracy vs depth plot. (c) Equivalent quantum circuit for classification of Iris dataset.}
\end{figure}

\begin{figure*}[htbp]
\centering
\includegraphics[width=\textwidth]{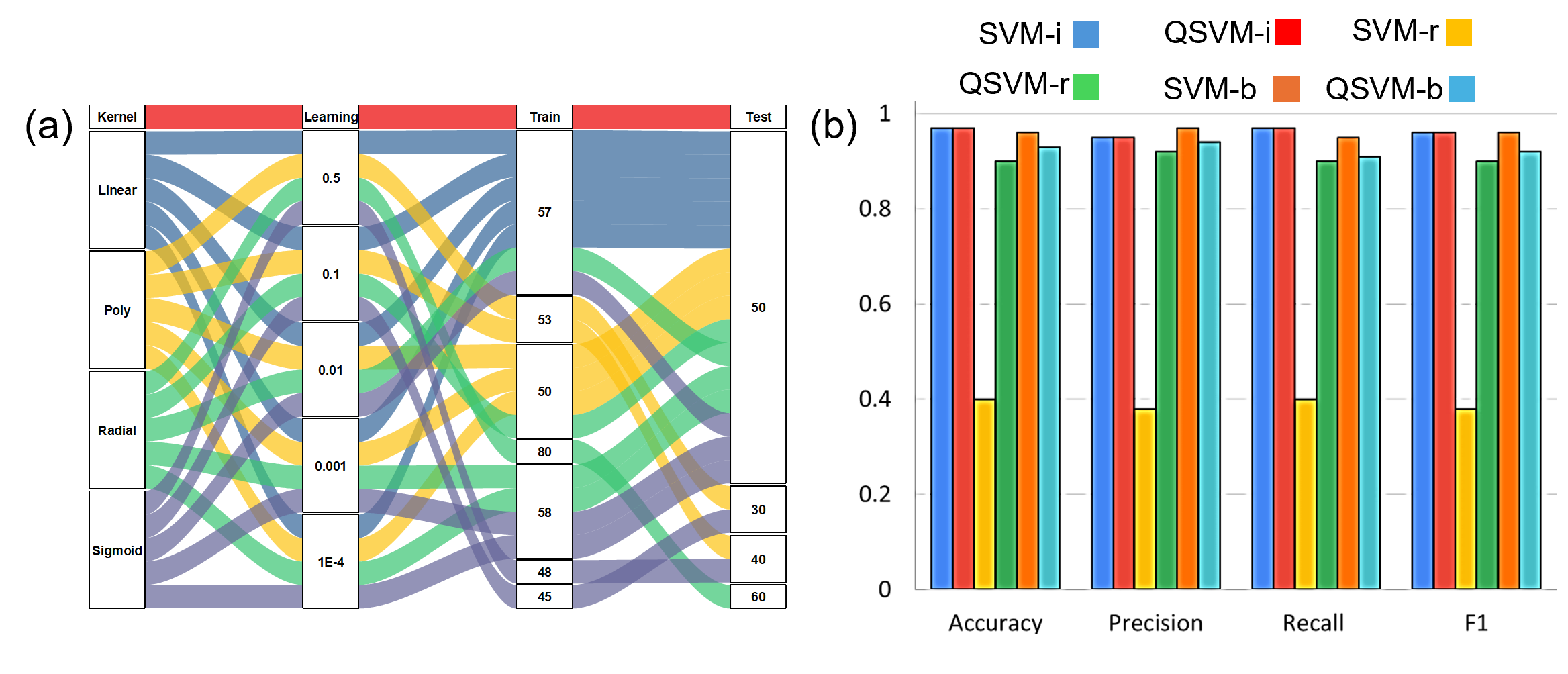}
\caption{(a) Different kernels testing for support vector machine (SVM). (b) The summary table of the different evaluation metrics.}
\end{figure*}

Understanding the specific operations and interactions within this circuit is crucial for interpreting its predictions and ensuring its transparency and reliability in medical applications. \cite{Ding2022, Moradi2022} The potential of combining kernel methods and quantum computing for breast cancer diagnosis is further supported by the literature. 

To extend our investigation to more than binary classes, we have utilized the Iris dataset having 3 classes. Figure 3a presents a scatter plot that depicts instances of a dataset with features related to classification. However, the lack of clear labeling obscures the specific attributes used for classification, making precise interpretations difficult to find. The quantum circuit learning journey shows an increase in the training accuracy as the depth of the circuit increases, indicating its ability to grasp more refined patterns in the data (Figure 3b). The increased accuracy in Figure 3b prompts further investigation to confirm whether it indicates successful information extraction or potential overfitting. Quantum circuit, emphasizing the importance of understanding its functionality, specific gates, and connections to interpret the results and discern the potential advantages of this approach. The 4-qubit quantum circuit consists of a 4-hadammard gate to create the entanglement (Figure 3c). Additionally, the ability of quantum circuits to uniformly address the Hilbert space has been linked to classification accuracy, emphasizing the relevance of quantum computing in machine learning tasks. \cite{Li2015, Shan2022}

 The experiment depicted in the image explores the impact of different kernel types and learning rates on the performance of a machine learning model (Figure 4a). The study involved the use of a linear kernel, a polynomial kernel, a radial basis function (RBF) kernel, and a sigmoid kernel, with variations in the learning rate for each kernel. Performance evaluation was performed on both training and test data. The findings revealed that the choice of kernel and learning rate significantly influences model performance. For example, the RBF kernel with a learning rate of 0.01 exhibited the highest accuracy of 80\% on the training data but the lowest accuracy of 30\% on the testing data, indicating potential overfitting. 
In contrast, the linear kernel with a learning rate of 0.5 achieved the best performance on the test data with an accuracy of 60\%, suggesting a better generalization to unseen data. However, it showed a lower accuracy of 57\% in the training data, indicating potential underfitting. The other kernels yielded mixed results, with the polynomial kernel achieving 53\% precision on the training data and 50\% in the testing data, and the sigmoid kernel achieving 48\% accuracy on the training data and 40\% on the testing data. These results underscore the critical importance of carefully selecting the kernel and learning rate for machine learning models. 
The evaluation metrics for training the classical and quantum algorithms have indicated that there is not much deviation in accuracy when training the linear data, while for the random datasets, quantum machine learning performs better with higher accuracy 45\% (Figure 4b). It indicates the different evaluation metrics such as precision, precision, recall, and F1 score to compare the performance among different databases, where i indicates the iris data set, r indicates the randomly generated respiratory dataset, while b represents the breast cancer data set. The optimal choice depends on the specific problem and the data set, emphasizing the need for experimentation to identify the best combination for a given task. \cite{Elmaghraby2020, Kumar2023, Saxena2023, Neill2018}

\section{\label{sec:level1}Conclusion}

Classical SVMs often struggle with complex and randomly distributed datasets, compromising their accuracy and efficiency. Our proposed KQ-SVM leverages quantum-enhanced kernel functions and quantum parallelism to address these challenges. Empirical analysis across diverse datasets shows KQ-SVM significantly outperforms classical SVMs, achieving over 45\% higher accuracy on complex datasets while maintaining comparable performance on linear datasets. This research demonstrates the transformative potential of quantum computing in machine learning, paving the way for enhanced performance and accuracy in real-world applications.
\vspace{-1em}
\section{\label{sec:level1}Experimental Section}
\vspace{-1em}
\subsection{\label{sec:level2}Fabrication and Characterization of the Sensor}
The nylon-11 nanofibers were produced using the electrospinning technique. The PVDF solution has taken 10 wt\% in the mixed solution of Trifluoroacetic acid: Acetone in the ratio of 6:4 and heated up at 60 °C for 6 hours. Further solution was loaded into a 10 ml syringe while applying the 18 kV voltage on the syringe tip, and the produced nanofibers were collected on the rotating drum collector at 1200 rpm. The nanofiber mat is sandwiched between the aluminum electrodes to fabricate the piezoelectric sensor.  

\vspace{-1.8em}
\subsection{\label{sec:level2}Characterization Techniques}
A digital storage oscilloscope (DSOX1102G, Keysight) was used to acquire the open-circuit voltage and respiratory signals. All the measurements were acquired in the non-invasive mode on the author himself and volunteers. Written consent has also been given prior to data recording. 
\vspace{-1.8em}
\subsection{\label{sec:level2}Development of Machine Learning and Quantum Machine Algorithms}
\subsubsection{Quantum Machine Learning Algorithms}
We utilized the panda's library for data manipulation and sci-kit-learn for data preprocessing tasks such as feature scaling, dimensionality reduction, and train-test splitting. The Wisconsin Breast Cancer (Diagnostic) dataset is obtained from the UCI Machine Learning Repository.
We used Qiskit for quantum computing functionalities for the quantum-based classification model, Qiskit Machine Learning for implementing quantum kernels and QSVC, and Qiskit Algorithms for algorithmic support. Additionally, we employed sci-kit-learn for traditional machine learning models.

Our quantum-based model consisted of feature mapping using the ZZFeatureMap from Qiskit's circuit library, a fidelity quantum kernel implemented using the Fidelity Quantum Kernel from Qiskit Machine Learning, and the QSVC model for training and classification tasks. For comparison, we trained a classical Support Vector Classifier (SVC) using scikit-learn. We initialized the experiment with a fixed random seed for reproducibility and split the data set into training and testing sets with an 80:20 ratio using stratified sampling. All the quantum measurements have been carried out on the IBM computing platform.

\subsubsection{Classical Machine Learning Algorithms}
Support vector machine (SVM) has been built using Python libraries, including (NumPy, TensorFlow, and Matplotlib), and sequential data for a classification task. We loaded data into separate X and Y data frames, performed one hot encoding on the labels, and split the data set into 80\% training and 20\% testing. It was trained for 10 epochs with batch size 32, using categorical cross-entropy loss and the Adam optimizer. We evaluated model performance in the test set, visualized results with a confusion matrix using Seaborn, and tracked training accuracy over epochs.

\begin{acknowledgments}
The authors express their gratitude to the entire Quantum Accelerated Computing workshop team. AB extends appreciation to the University Grants Commission (UGC) for the fellowship (1354/(CSIR-UGC NET DEC. 2018)). The authors deeply appreciate Param Smriti for providing the high-performance computing facility essential for conducting this work. Furthermore, we are sincerely grateful for the support from the IBM quantum computing facility and Pennylane for their contributions to quantum computing resources. The authors are very grateful for the fruitful discussion with Dr. Gurumohan Singh, CDAC-Mohali.
\end{acknowledgments}

\nocite{*}

\bibliography{apssamp}% Produces the bibliography via BibTeX.

\end{document}